# Evolution of biosequence search algorithms: a brief survey

**Gregory Kucherov**

CNRS and LIGM/University of Paris-Est

Marne-la-Vallée, France

`Gregory.Kucherov@univ-mlv.fr`

**Abstract** The paper surveys the evolution of main algorithmic techniques to compare and search biological sequences. We highlight key algorithmic ideas emerged in response to several interconnected factors: shifts of biological analytical paradigm, advent of new sequencing technologies, and a substantial increase in size of the available data. We discuss the expansion of alignment-free techniques coming to replace alignment-based algorithms in large-scale analyses. We further emphasize recently emerged and growing applications of sketching methods which support comparison of massive datasets, such as metagenomics samples. Finally, we focus on the transition to population genomics and outline associated algorithmic challenges.

## Introduction

The rise of bioinformatics in the 90s as a major application field of computer science was powered by mass obtention of sequential composition of biological macromolecules: DNA, RNA, proteins. Faithfully representing those molecules as strings over a small fixed alphabet coupled with available technologies for obtaining such strings provided unprecedented opportunities for computer scientists to promptly apply their algorithmic ideas to biological analyses. Even if many other types of data are produced nowadays by genomic and proteomic technologies (HiC, mass spectrometry, microarrays, ...), biosequence data remains at the core of bioinformatics.

Since the beginning of nucleotide sequencing, genomic sequence data has been produced at an ever-increasing pace. However, this growth has not been steady as it was marked by several major innovations of sequencing technology. Each of them called for new algorithmic techniques to meet new computational challenges.

For the last years, genome sequencing became a major generator of big data which is quickly catching up most prominent big data applications -- astronomy or social media [1]. The *Sequence Read Archive* (http://www.ncbi.nlm.nih.gov/Traces/sra/) has grown explosively since 2007 and currently contains about 15 peta-bases with almost half of it in open access. The total amount of produced sequence data is currently estimated to double every seven months [1], which is to be contrasted with processor performance doubling every two years, according to the famous Moore's law. Today's genomic sequencing projects often involve huge, sometimes terabyte-sized datasets [1], in contrast to about 25GB of data used for sequencing human genome in 2001 [2], [3]. This evolution calls for a new revision of algorithmic foundations of DNA search and comparison.

Distributed and cloud computing can provide infrastructures for storing and processing large data and avoiding costly data transfers. However, these technologies, if available, can only be a part of the solution and must be complemented by novel software tools [4]. This, in turn, requires the development of novel computational techniques based on new algorithmic methods and efficient data structures.

Another avenue in analyzing big sequence data is associated with the recent explosive interest in machine learning techniques, in particular in deep learning, inspired by spectacular results in image analysis, speech recognition and some other areas. This development did not avoid bioinformatics: for the last few years, a number of works have been published applying deep learning to a range of bioinformatics problems [5]. While this direction is promising and is likely to be continued in the coming years, deep learning often requires massive training datasets and a powerful computer infrastructure as well. Due to their interpretability and controlled precision, "traditional" algorithmic techniques remain predominant in biological sequence analysis.

The goal of this article is to overview key algorithmic ideas introduced in response to the evolution of DNA sequencing technologies. We specifically focus on most recent developments and highlight algorithmic techniques that are studied in order to face the current tremendous inflow of DNA sequence data.

### Early days: dynamic programming

As biological sequences were becoming increasingly available starting from the 80s, detecting similarities between them opened up various new opportunities for biological studies. The distance measure on biosequences is dictated by statistics of mutational transformations that these sequences undergo in the course of evolution. The usual model for comparing relatively short sequences (such as genes) only considers *point mutations* i.e. character substitutions, insertions or deletions. However, the simple edit distance is not good enough to reflect biological proximity and is then replaced by *weighted edit distance* where different character substitutions may get different penalties [6]. Comparing two sequences in their entirety is then formalized as the *optimal global alignment* problem and is solved by the classical dynamic programming algorithm known in bioinformatics as the Needleman-Wunsch algorithm [7].

The Smith-Waterman algorithm [8] is a modification of Needleman-Wunsch for computing an *optimal local alignment*, i.e. the best global alignment of *substrings* of two input sequences. This computation is relevant to a wider range of biological tasks that involve searching for similar fragments across sequences, such as similar genes between genomes or similar protein domains between proteins. Gotoh's algorithm [9] further extends Needleman-Wunsch and Smith-Waterman to the case of *affine gap penalties* when the penalty for $k$ consecutive insertions (or deletions) is the sum of a *gap opening penalty* and $k$ *gap extension penalties*.

All these algorithms, however, have time complexity $O(nm)$ for $n,m$ being the lengths of involved sequences, and therefore are too computationally demanding for massive sequence search.

### Filtration-based heuristics and database search

By the end of the 80s, the total volume of available DNA sequencing data reached the scale of tens of millions of nucleotides and began to be organized in databases that could then be searched for sequence queries. This resulted in the emergence of a new family of tools most prominently represented by the popular BLAST software [10], [11]. From the algorithmic viewpoint, these tools implement a heuristic approach to sequence alignment guided by two general ideas: *filtration* and *indexing*.

Filtration consists here in identifying those sequence fragments of the database which share with the query small matching patterns called *seeds*. The rest of the database is then discarded from consideration ("filtered out") and seeds are attempted to be extended into meaningful alignments using a costlier algorithm, possibly involving dynamic programming. The search uses a pre-built index of the database, typically a hash table, speeding up location of seed patterns found in the query. The performance of this approach is defined as a trade-off between the fraction of detected alignments under interest (sensitivity) and the rate of "false positive" spurious seeds (specificity) and is determined by the type of seeds used for filtration. For example, an early version of BLAST for DNA sequences (blastn) uses the default of 11 consecutive matching nucleotides shared with the query to trigger the extension step.

While the popularity of BLAST made of it a ubiquitous tool that is still used nowadays for small-scale sequence comparisons, an elegant algorithmic improvement to BLAST was proposed around 2002: it was shown that the sensitivity-specificity trade-off can be improved "for free" by replacing contiguous seeds ($k$ consecutive matching nucleotides) by $k$ matching nucleotides sampled according to a given pattern, called *spaced seed*. This seemingly insignificant modification turned out to bring a remarkable improvement due to statistical properties of patterns in random sequences [12]. The superiority of spaced seeds applied to DNA search was first established in [12], [13] in relation to PatternHunter software and was later exploited by many other algorithms (e.g. [14], [15]). Remarkably, the choice of a specific spaced seed can be optimized for the class of sequences under consideration [16], further improving flexibility and adaptability of the search.

BLAST remained a "Swiss army knife" for bioinformatics analyses for almost two decades that prepared the age of whole genome sequencing. Starting with the sequencing of several first model organisms in the late 90s, the human genome sequence was published in 2001 (draft [2], [3]) and 2004 (completed version [17]), marking the "end of the beginning" (using the expression of [18]) of the *genomics* era characterized by the change of the methodological focus in biological studies from individual genes to whole genomes.

## Burrows-Wheeler transform and NGS revolution

The mid-2000s saw a major technological shift due to the advent of so-called New Generation Sequencing (NGS) technologies. These sequencers produced very short DNA fragments called *reads* (35 to 400 nucleotides), in huge quantities (tens of GB of sequence data per run) and at a low cost [19], [20]. This development made a tremendous impact to numerous fields of genomic research by making possible genome-wide or population-wide studies that were previously unfeasible or simply impossible to carry out. Applications of NGS are not limited to "reading" genomic DNA, other major application areas include sequencing RNA transcripts (RNA-seq), analyzing DNA methylation (Methyl-seq), studying protein-DNA interactions (ChIP-seq) and proteomic profiles (exome-sequencing). We refer to [20] for a recent panorama of NGS techniques and applications.

The most common task in processing NGS sequencing data is to locate reads within a reference genomic sequence. Here, input reads typically come from an organism under study (e.g. a diseased individual) and the reference genome is a known representative genomic sequence of the species (e.g reference human genome sequence), or possibly a phylogenetically related species. This task, known as *read mapping*, amounts to aligning reads against a reference sequence. A major computational difficulty, however, comes from the size of data: millions of reads have to be aligned to a sequence that can contain billions of letters. BLAST-like tools simply cannot cope with such data in a reasonable time. As an example, mapping a modest million of 100bp-reads to the average-size human chromosome 9 takes BLAST more than 24 hours on a powerful 12-core computer. This led to the development of a number of specialized alignment tools for read mapping.

A huge variety of read mapping software have been developed, see e.g. Sect 3.1.4 of [21]. Most common practical mappers include BWA-MEM [22], Bowtie 2 [23], GEM [24] that are several orders of magnitude faster than BLAST-like tools [4]. Thus, BWA-MEM spends about 6 minutes to accomplish the above-mentioned mapping task. The source of efficiency of these algorithms is the indexing structure known as BWT-index or FM-index [25].

BWT-index takes its roots in word combinatorics: it is based on the Burrows-Wheeler transform [26] which, from the combinatorial perspective, is closely related to the Gessel-Reutenauer bijection [27]. While the primary application of BWT was text compression [28], it has been applied to construct a compact text index [29] that turned out to be surprisingly powerful.

As opposed to hash-based indexes underpinning seed-and-extend strategies represented by BLAST, BWT-index is a *full-text index* as it supports search for arbitrary-size strings. BWT-index belongs to a more general family of *succinct indexes* whose size *in bits* is close to the information-theoretic minimum needed for the lossless representation of the object (here the sequence) [30]. In practice, BWT-indexes for DNA data take a few bits per character, e.g. popular mappers such as BWA-MEM or Bowtie2 index both the input sequence and its complement using less than two bytes per character. This is an order of magnitude less compared to classical full-text indexes such as suffix trees or suffix arrays. Interestingly, although much more compact, a BWT-index can provide even more functionalities than classical indexes: for example, BWT-index can be extended to support *bi-directional search* of patterns [31] which enables more efficient algorithmic solutions [32], [33].

In sum, with the rise of NGS technologies, sizes of DNA datasets shifted from the megabyte to gigabyte scale. BWT-index can be considered the main novel algorithmic tool that came in response to this transition. As a result, generic BLAST-like tools gave way to dedicated NGS read alignment software, mostly based on BWT-index.

## Second life of alignment-free methods

The sheer size of modern datasets often makes it hard to apply even most efficient sequence alignment software. One illustration to this is *metagenomics* where the task may consist in mapping millions of reads to thousands of microbial genomes in order to elucidate the composition of an environmental sample under study [34]. Even specialized mapping software may not meet practical time requirements for this task. On the other hand, in such applications as metagenomics, computing alignments may not be indispensable as we often seek to only identify the genomes the given read could originate from.

A way to cope with that is to abandon the idea of computing alignments and replace it by a weaker but easier-to-compute similarity measures to allow for speed-ups and memory gain at the price of lower accuracy. A common idea is to view a sequence as a (multi-)set of *patterns* that occur in it. In the simplest case, those patterns are substrings of fixed length $k$ ($k$-mers). Then, similarity between sequences is expressed through the similarity between corresponding sets. This approach is known as *alignment-free* or *composition-based comparison*.

While alignment-free methods have been previously considered in bioinformatics [35], their application was not common until recently, as most of the efforts have been put on improving alignment algorithms. On the other hand, alignment-free methods have been extensively and for a long time explored in machine learning [36] or information retrieval [37], however classical machine learning techniques (such as kernel-based methods for example) are in general hardly feasible at the scale required by modern bioinformatics applications.

Today, new resource-frugal alignment-free techniques are rapidly coming into broad use. To continue with the example of metagenomics, most of modern metagenomic classifiers for whole-genome sequencing data are based on *k*-mer analysis [38]. This development raises new algorithmic questions. Scaling to required data sizes imposes computationally lightweight algorithms together with extremely memory-efficient optimized data structures.

Besides the above-mentioned BWT-index, several proposed data structures build on *Bloom filters* -- a classic data structure for memory-demanding applications. These extensions include Cascading Bloom Filter [39], Bloom Filter Trie [40], Sequence Bloom tree [41] and its variants [42]. Space-efficient and cache-friendly lossless hashing schemes, such as *counting quotient filters*, have been proposed as an alternative to Bloom filters [43].

Besides coping with memory constraints, a general algorithmic challenge in this class of methods is to achieve the best accuracy in capturing sequence similarity under very strong time and memory restrictions. For example, in metagenomic classification of short Illumina(TM) reads, *k*-mer-based methods may already achieve an accuracy comparable to Blast-like methods [44]. Interestingly, the accuracy can be improved by replacing *k*-mers by spaced *k*-mers [45][46], in the same vein as it was done for computing alignments.

## Sketch-based methods

Tera-base scale datasets, increasingly common in modern applications, impose new constraints as they inherently cannot be indexed within GB-scale memory without sacrificing the exhaustive character of the index. In this context, the last few years saw a growing interest in methods based on *locality-sensitive hashing* (LSH). In a nutshell, the idea of LSH is to map complex objects (e.g. long sequences) to smaller objects, called *sketches* or *fingerprints*, such that similar objects are likely to be mapped to similar or identical sketches. This allows one to replace operations on input objects (search, comparison, sequence overlap, ...) by corresponding operations on sketches.

Starting from the seminal paper [47], the concept of LSH has been extensively studied from the theoretic perspective [48]. On the other hand, a specific instance of LSH, called *MinHash*, was applied earlier to finding similar web documents [49]. MinHash can be seen as following the alignment-free paradigm discussed earlier, as it considers sequences as sets of their *k*-mers (originally called *shingles*) and measures sequence similarity as the Jaccard index on those sets. The Jaccard index is estimated by comparing sketches of the sets defined using minimum hash values of corresponding *k*-mers under one or several hash functions.

MinHashing, in its basic form, only works for comparing datasets of similar size. A way to cope with difference in size is to use *winnowing* [50], that is to compute a MinHash for each sequence window of a fixed size *w*. If only one hash is computed for each window, this leads to the concept of *minimizer* [51]: a minimizer is a minimum (under a given order, e.g. lexicographical) *k*-mer within some surrounding window of size *w*.

Several recent works employed MinHashing for performing various very large-scale analyses of sequencing data. In [52], the authors propose a formula to transform the MinHash estimation of the Jaccard index into a more biologically meaningful parameter: nucleotide identity rate of compared sequences. Based on this measure, they were able to conduct several large-scale computations, such as sketching and clustering the entire RefSeq genome database (release 70) [53] amounting to over 600GB of sequence data.

Besides dealing with very large datasets, MinHashing provides a way to improve seed-based sequence search (cf above) by using minimizers as seeds. This idea is particularly useful for mapping long reads produced by "third-generation sequencing technologies", such as Pacific Biosciences(TM) or Oxford Nanopore(TM), presenting high error rates dominated by insertions/deletions (indels) of short fragments. Here, minimizers can act as "marker fragments" supporting search algorithms dealing with indels in a more flexible way. This direction has been followed by recently proposed mappers for long reads [54], [55].

Another deficiency of basic MinHash is its inability to deal with multiplicities of data items, such as $k$-mer frequencies for example. This, however, can be handled by *Consistent Weighted Sampling* [56] (or *Weighted MinHash* [57]) – a generalization of MinHash to multiplicity vectors. This technique was recently applied to the comparison of microbiome samples [58].

Computationally frugal techniques based on Bloom filters or sketching enable efficient queries to very large corpora of sequence data. In a way similar to web search, such *search engines* make it possible to perform bioinformatics analyses in a much faster and simpler way. One of most recent works illustrating this approach [59] reports building a Bloom-filter-based index of all known genomes of bacteria and viruses (170TB overall) making this data available for sequence search queries. Thus, the prevalence of a certain gene, for example, can be evaluated within a single search query. Many other applications of search engines can be thought of, such as looking up a gene transcript in an archive of RNA-Seq experiments [41].

## From genomics to population genomics

Today, we are living in a time of extensive sequencing of individual genomes of various species. This, above all, applies to humans. Many countries have undertaken nation-wide projects of genotyping their populations [60]. A prominent example here is Iceland that, by today, collected genotyping data for at least half of its citizens, of which tens of thousands have their whole genome sequences established [61]. The *Genome of the Netherlands* (http://www.nlgenome.nl/) project aims at compiling comprehensive maps of genomic haplotypes through massive sequencing on a whole country scale, and other countries adopted similar programs. Altogether, according to some estimations, 100 million to 2 billion of individual human genomes will be sequenced in the world by 2025 [1].

Genotyping multiple genomes is not restricted to the study of inter-individual genetic variability but includes intra-individual heterogeneity within a population of cells as well. A notable example is provided by cancer cell genomes that show extensive heterogeneity among somatic mutations, both intra- [62] and inter-patient [63]. Single-cell sequencing technologies enable sequencing specific cancer subclones [64]. Large international consortia currently collect enormous amount of data about genomic changes in different cancers. Among them, *Genomic Data Commons (GDC)* by National Cancer Institute (https://gdc.cancer.gov/) hosts several petabytes of sequence or alignment data.

Finally, multiple genomes are being sequenced for many species other than human as well. Certain bacteria, especially important human pathogens, have up to tens of thousands sequenced strains (cf e.g. [65]). Genomic diversity of various plants is being actively studied as well. For example, hundreds of rice accessions have been sequenced [66]. Some other crop lines (maize, tomato, barley, ...) as well as other plants (e.g. *Arabidopsis thaliana*) are being actively studied.

Supported by this mass sequencing, genomics is now entering a new phase characterized by a new shift of the underlying methodological paradigm: the basic object of genomics is now switching from a "reference genome" of a species to a collective genome of a population of individuals or cells, sometimes called a *pan-genome* [67]. This transition is comparable to the shift "from genes to genomes" (or "from genetics to genomics") occurred at the beginning of massive DNA sequencing in the late 90s, resulting in the change of the methodological focus from individual genes to whole genomes.

Seen from the computational perspective, the main question of population genomics is how to represent a collective genome of a population to allow an efficient algorithmic processing and, on the other hand, to capture variations in individual genomes. One crucial factor here is a limited genetic variability within a species: e.g. only about 0.1% of the genome differs between two human individuals. This suggests an approach when common genomic fragments are represented in a single copy, leading to replacing a single reference genomic sequence by a labelled graph whose paths represent individual genomes [68], [69]. This representation, however, has its limits as it does not capture interactions between different parts of a genome (such as epistatic interactions between remotely located genes) unless additional labeling of graph paths is stored.

*Compressive genomics* [70], [71] is a paradigm that tries to benefit from the global "topological" structure of genomic data. On the one hand, biological sequences of the same or close species are highly similar. On the other hand, close sequences are evolutionary related in general, and, as a consequence, the space of all existing genomic sequences has low *fractal dimension* [71]. This suggests that sequences can be stored in a compressed form and search can be done by first locating the vicinity of target sequences without prior decompression, and then by refining the search within this vicinity. Proof-of-concept experiments confirm the soundness of this approach [70], however practical search engines based on this paradigm are still to prove feasible.

Coping with the exponential growth of genomic data would have an immense impact on virtually the whole genetic research, ranging from elucidating fundamental genetic mechanisms to providing new keys to understanding numerous genetic diseases. To give just one illustration, our ability to grasp, represent and computationally approach a cancer pan-genome appears a necessary step on the way to new cancer therapies. We consider that meeting this challenge will be a major challenge on the bioinformatics agenda for the years to come.

### Acknowledgements

Many thanks go to Karel Břinda for his helpful comments and suggestions.